\documentclass[twocolumn,%
nofootinbib,preprintnumbers
]{revtex4}
\preprint{TUW-04-07}
\preprint{YITP-SB-04-10}
\preprint{ITP-UH-08/04}
\usepackage{graphicx}
\usepackage{latexsym}
\usepackage{amssymb}

\newcommand{\be}{\begin{equation}}
\newcommand{\ee}{\end{equation}}
\newcommand{\bea}{\begin{eqnarray*}}
\newcommand{\eea}{\end{eqnarray*}}

\newcommand{\SUSY}{susy}

\def\g{\gamma}

\def\l{\lambda}
\def\m{\mu}

\def\s{\sigma}
\def\t{\tau}

\def\z{\zeta}
\def\0{\over } \def\1{\vec } \def\2{{1\over2}} \def\4{{1\over4}}
\def\5{\bar } 
\def\6{\partial }
\def\7#1{{#1}\llap{/}}
\def\8#1{{\textstyle{#1}}} \def\9#1{{\bf{#1}}}
\def\.{\dot }
\def\^#1{\widehat{#1}}

  \let\g=\gamma 
 \let\z=\zeta  
  \let\l=\lambda \let\m=\mu
    \let\s=\sigma
\let\t=\tau  

\let\ph=\varphi

\def\CH{{\cal H}}

\def\CL{{\cal L}}

\def\({\left(} \def\){\right)} \def\<{\left\langle } \def\>{\right\rangle }
\def\[{\left[} \def\]{\right]}  

\def\pmbf#1{\setbox0=\hbox{${#1}$}
        \kern-.025em\copy0\kern-\wd0
        \kern.05em\copy0\kern-\wd0
        \kern-.025em\raise.0433em\box0 }

\def\be{\begin{equation}}
\def\ee{\end{equation}}

\newcommand{\bel}[1]{\begin{equation}\label{#1}}
\def\bea{\begin{eqnarray}}
\newcommand{\beal}[1]{\begin{eqnarray}\label{#1}}
\def\eea{\end{eqnarray}}
\def\nn{\nonumber\\ }
\def\eqa{\begin{eqnarray}}
\def\eqae{\end{eqnarray}}

\newcommand{\ds}{\!\not\!\partial}

\newcommand{\unit}{1\!\!1}
\newcommand{\intsum}{\sum\!\!\!\!\!\!\!\int}

\newcommand{\pv}{\phi_{\mathrm V}}
\begin{document}
\title{New developments in
the quantization of supersymmetric solitons\\
(kinks, vortices and monopoles)
%
}

\author{Anton Rebhan}
\affiliation{Institut f\"ur Theoretische Physik, Technische
Universit\"at Wien,\\Wiedner Haupstr.~8-10, 
A-1040 Vienna, Austria }

\author{Peter van
Nieuwenhuizen}
\affiliation{C.N.Yang Institute for Theoretical
Physics, State University of New York,\\ Stony Brook, NY 11794-3840, USA }

\author{Robert Wimmer}
\affiliation{Institut f\"ur Theoretische Physik,
Universit\"at Hannover, \\Appelstr.~2, 
D-30167 Hannover, Germany}

\begin{abstract}
We discuss the one-loop quantum corrections to
the mass $M$ and central charge $Z$ of supersymmetric (susy)
solitons: the kink, the vortex and the monopole. 
Contrary to previous expectations and published results, in each of these
cases there are nonvanishing quantum corrections to
the mass. For the $N=1$ kink and the $N=2$ monopole 
a new anomaly in $Z$ 
rescues BPS saturation ($M=Z$);
for the $N=2$ vortex, BPS saturation is rescued for two reasons: (i) the
quantum fluctuations of the Higgs field acquire a nontrivial
phase due to the winding of the classical solution, and (ii) a
fermionic zero mode used in the literature is shown not to be
normalizable.
\end{abstract}

\maketitle

\section{An introduction to 
new developments in quantum field theory for susy solitons}

Solitons \cite{Raj:Sol} have recently come back in the center of attention of
quantum field theory (QFT) because in a certain class of
supersymmetric (susy) field theories (the ones with $N=2$ susy)
dualities between field theories with point-particles and field
theories with solitons allowed calculation of some
nonperturbative effects~\cite{Seiberg}. 
Let us begin by defining
what we mean by a soliton:
\begin{quote}
{\em Definition 1:} a soliton is a nonsingular time-independent
solution of the classical field equations in {\bf Minkowski}
spacetime with finite {\bf energy}. 
\end{quote}
Solitons can be viewed as
extended particles (``lumps'') which should clearly have finite
mass (= finite energy at rest). We shall only consider
relativistic field theories. There exist also time-dependent
solutions with finite energy (the ``breather'' solution in the
sine-Gordon model, for example), but we consider only
time-independent solitons.  One can, of course, boost solitons in
a relativistic theory, and obtain then moving solitons, but since
one can always choose a Lorentz frame in which they are at rest,
we restrict our attention to only time-independent solutions.

A soliton is closely related to an instanton. The later is defined as follows:
\begin{quote}
{\em Definition 2:} an instanton is a nonsingular solution of the
classical field equations in {\bf Euclidean} space with finite
{\bf action}.
\end{quote}
Because instantons have finite action, they
contribute already at the classical level to the path integral.
This is the reason for the requirement of finite action. It is
clear that a soliton in $n+1$ dimensions is an instanton in $n$
dimensions: since the time coordinate plays no role in the
soliton solutions, the space integral in Minkowski space can also
be viewed as an integral over Euclidean space, and the energy
$E=\int d^n x {\cal H}(x)=\int d^nx(p\dot{q}-{\cal L})$ is equal
to the action $S=-\int d^nx{\cal L}$.

We shall discuss three solitons:
\begin{itemize}
\item[1)] the {\bf kink} in 1+1 dimensions with $N=1$ susy
\item[2)] the {\bf vortex} in 2+1 dimensions with $N=2$ susy
\item[3)] the {\bf monopole} in 3+1 dimensions with $N=2$ susy
\end{itemize}
There exist also susy extensions of these solitons with more susy
($N=2$ for the kink, $N=4$ for the monopole) or less susy ($N=1$ for the vortex). In
addition, discussions have been given about the quantum
corrections to these solitons without fermions and without susy.

All three models have recently led to surprising new insights in the
quantization of solitons. In particular, a new anomaly was discovered
in the quantum corrections to the central charge of the susy algebras
for these models. The solitons have not only a mass $M$ but also a
central charge $Z$ \cite{Witten:1978mh}.  
The quantum mass of a soliton is obtained by
evaluating the vacuum expectation value of the Hamiltonian.  The
latter depends on the soliton background because one decomposes
quantum fields $\phi$ into a background field and a quantum field.
The quantum fluctuations appear quadratically 
(and of course also at higher orders) in the action, and thus
the quantum correction to the mass is to first order in $\hbar$ the
sum over the zero point energies of all the fluctuations.  The quantum
central charge arises as follows.  The susy generators are the Noether
charges for rigid susy, and as any Noether charge they are the space
integrals over the time components of the Noether currents.  They are
expressed in terms of the Heisenberg fields $\phi$.  Using equal-time
canonical (anti) commutation relations, one finds then for example for
the kink at the full quantum level before regularization
\eqa && \{ Q^\pm , j^\pm (x) \} =
\CH (x) \pm \z (x) \nn && j^\pm = \dot\phi \psi_\pm + (\partial_x \phi
\pm U) \psi_\pm \nn && Q_\pm = \int j_\pm (x) dx \nn && \z (x) =
\partial_x \phi(x) U (x) = \partial_\s W (\phi) \eqae where $U$ is a
potential quadratic in $\phi$ and $W (\phi) = \int U (\phi) d \phi$.

Classically, $M=Z$, the
well-known BPS bound, but quantum mechanically both $M$ and $Z$
get quantum corrections. At first sight (or thought) it is
surprising that $Z$ gets any quantum corrections at all, because
classically $Z$ is the space integral of a total space-derivative.
So it gets its contributions from far away from the soliton, and
if one can neglect the presence of the soliton, how can one still
get nonvanishing corrections? The answer (one answer) is that the
central charge is a composite operator, and applying point
splitting as a regularization scheme, the total derivative ceases
to be a total derivative, and one gets then quantum corrections
also from the region where the soliton is.

We shall not use point splitting as regularization scheme, but dimensional regularization.
There are actually two versions of dimensional regularization
\begin{itemize}
\item[(1)] ordinary ('t Hooft-Veltman-Bollini-Giambiagi) dimensional regularization \cite{'tHooft:1972fi,Bollini} where
$n$ becomes larger than $n_0=2,3$ or 4 for our models
\item[(2)] dimensional reduction \cite{Siegel}, 
where one lets $n$ get smaller than $n_0$. Vector
fields $A_\mu$ decompose then into $n$-dimensional vector fields 
with $0\leq\mu\leq n$ and scalar fields (called
$\epsilon$-scalars) where $n\leq\mu\leq n_0$ (and
$n_0-n=\epsilon$).
\end{itemize}

If $n>n_0$ one needs a model which remains susy in higher
dimensions if we want to preserve susy at the regu\-larized level.
 This is in general not possible because the number of components
of spinors grows much faster in higher dimensions than that of
bosons. However, in the low dimensions we consider there are
possibilities. For example, the $N=1$ (more precisely $N=(1,1)$)
susy kink in 1+1 dimensions
remains $N=1$ susy in 2+1 dimensions because spinors have 2
components both in 2 and in 3 dimensions. For the vortex we
preserve $N=1$ susy in 3+1 dimensions by starting with an $N=2$
model in 2+1 dimensions ($2\times 2= 1\times 4$). Finally, for the
monopole we get a minimal susy model in 5+1 dimensions (with
complex unconstrained 4-component chiral spinors) by starting with
an $N=2$ model in 3+1 dimensions with two real 4-component spinors
(or, equivalently, two complex 2-component spinors).

If $n<n_0$, the models remain automatically susy because one does
not change the number of bosonic and fermionic field components
(one only reinterprets some components of vector fields as
$\epsilon$-scalars as we have discussed). So much for how we deal
with susy in the two versions of dimensional regularization. We
wish to preserve (ordinary, see below) susy, and the best way to
preserve it is to preserve it manifestly, by using models which
remain susy even after regularization.

How can one understand that an anomaly in $Z$ is present?
Anomalies form multiplets in susy, and these anomaly multiplets
contain the trace anomaly and the conformal susy anomaly
$\gamma\cdot j$. Here a small explanation is needed.
Supersymmetry is the same as translational symmetry into the
fermionic directions in superspace, and like ordinary
translational symmetry, there is no anomaly in ordinary susy:
$\partial^\mu j_\mu$(susy)=0 at the quantum level. However, in
massless theories one also has conformal susy (the fermionic
counterpart of scale invariance and conformal boost invariance),
and the  conformal susy current is $j_\mu ({\rm
conf})=(x^\rho\gamma_\rho)j_\mu({\rm susy})$. Clearly,
$\partial^\mu j_\mu ({\rm conf})=\gamma^\mu j_\mu ({\rm susy})$ if
ordinary susy is free from anomalies. In massless (conformal)
theories the ordinary susy current satisfies at the classical
level the relation $\gamma^\mu j_\mu ({\rm susy})=0$. For
example, in the WZ model in 1+1 dimensions, $j_\mu ({\rm
susy})=(\partial \!\!\!/ \varphi)\gamma_\mu\psi$, and since
$\gamma^\mu\gamma_\rho\gamma_\mu=0$ in 1+1 dimensions, one
has $\gamma\cdot j ({\rm susy})=0$. In 3+1 dimensions the WZ model
yields $j_\mu ({\rm
susy})=F_{\rho\sigma}\gamma^{\rho\sigma}\gamma_\mu\psi$, and since
$\gamma^\mu\gamma^{\rho\sigma}\gamma_\mu=0$ in 3+1
dimensions, also here $\gamma\cdot j ({\rm susy})=0$. At the
quantum level, there can be anomalies in $j_\mu ({\rm susy})$
(the current $j_\mu ({\rm conf})$ is no longer conserved,
equivalently $\gamma^\mu j_\mu ({\rm susy})$ is no longer
vanishing).

What is the technical reason that  there is an anomaly, how does
it appear when one calculates? Let us recall that an anomaly
usually appears as 0/0: 0 because classically there is no
violation, and 1/0 because a quantum field has infinitely many
degrees of freedom (there are no anomalies in quantum mechanics).
Consider now first ordinary dimensional regularization. It is
known that translations in higher dimensions become central
charges in lower dimensions. For example, the susy kink model has
3 translations in 3 dimensions, which become 2 translations and
one central charge if one performs dimensional reduction (setting
the coordinate $x^3$ to zero). This seems to doom the prospects of
anomalies in $Z$, because we already mentioned that there are no
anomalies in the translational symmetry. However, the presence of
the soliton in the background {\bf polarizes} the fermionic
excitations in the extra dimensions \cite{Callan,DaRold:2001na}. 
This phenomenon has the same
origin as the quantum Hall effect, and as a result the
left-moving and the right-moving modes of the fermions have
different normalizations on the domain wall created by the
soliton. One word about domain walls 
\cite{Bollini,deCarvalho:1986sc,Graham:2001dy,deCarvalho:2001da,Izquierdo:2003yc}: 
if one moves from 1+1 to
2+1 dimensions, with coordinates $(x,y,t)$ and $y$ is the new
coordinate, then the soliton $\varphi (x)$ remains a solution with
finite energy per unit of length in the extra dimension. In 1+1
dimensions the energy density of the soliton solution $\varphi (x)$ is
located around $x=0$, so in 2+1 dimensions it is located around
the ``wall" (line in this case) $x=0$ but any $y$. One must then
solve the Dirac equation in 2+1 dimensions, and in this way one
discovers the polarization. (In string theory, Horava and Witten
\cite{Horava:1996ma}, and
others have used this effect for other purposes\footnote{They
considered 2 branes in $d=11$ supergravity with a chiral spinor on
one brane and an antichiral spinor on the other (this combination
cannot be avoided because there are no chiral spinors in odd
dimensions). Then they sent one brane to infinity, applied
dimensional reduction, and found in this way chiral spinors in 4
dimensions.}). A detailed calculation reveals that for the kink
and the monopole there is indeed an anomaly coming from the
polarization of domain wall fermions. Two remarks should quickly
be made: (1) in odd dimensions there are no anomalies, so how $Z$
gets a contribution for the vortex should at this point not yet
be clear to the uninformed reader, and furthermore, (2) there
are also massless fermions on the domain wall; they are chiral
(they only move in one direction) but these fermions do not
contribute to the anomaly because in dimensional regularization
massless tadpole integrals vanish. Only the polarization of the
massive domain wall fermions yields an anomaly.

In the calculation of quantum corrections to $M$ and $Z$ one
should let $n$ tend to $n_0$, so $\epsilon=n_0-n$ tends to zero.
However the sum of all polarizations is divergent. As the reader
may be now anticipate one indeed finds a total correction of the
form $\epsilon / \epsilon$ which is finite. Many other
regularization schemes have been analyzed, and also in these
schemes one finds an anomaly in $Z$ and the value of this anomaly
is the same in all cases. 

The picture becomes now clear in even dimensions (kink,
monopole). There are nonvanishing quantum corrections to the mass,
partly due to nonanomalous corrections (which however
vanish for kink and monopole in the most widely used
renormalization scheme) and partly due to the trace
anomaly. There are also nonvanishing corrections to the central
charge, due to the central charge anomaly (which sits in the same
multiplet as the conformal susy anomaly and the trace anomaly).
Both corrections are equal: the classical BPS bound also holds at
the quantum level.

In odd dimensions, in particular for the vortex, there is another
reason why there is a nonvanishing correction to $Z$. There cannot 
be an anomaly, as we already mentioned, but now the soliton
deforms the quantum fluctuations of the scalar (Higgs) field such
that the latter acquires an extra phase. This phase has a
nontrivial space dependence (a dependence on the angles which can
not be removed continuously by a gauge transformation which is
everywhere regular). This twist remains far away from the vortex
and does give a nonvanishing contribution to the integral of the
total derivative. The result is $M=Z$ also at the one loop level.

One would like to know whether there is an explanation for this
equality. This involves fermionic zero modes, let us first define
them.
\begin{quote}
{\em Definition 3:} A zero mode is a time-independent solution of the
linearized field equation for the quantum fluctuations which is
{\bf normalizable and nonsingular}. 
\end{quote}
Bosonic zero modes can be
obtained by making a symmetry transformation on the soliton, for
example shifting or rotating the classical solution. Many, but
not all, fermionic zero modes can be obtained by making an
ordinary susy transformation of the soliton. (For instantons there
exist fermionic zero modes for $SU(n)$ with $n\geq 3$ which do
not come from susy).

Arguments have been given in the literature that the equality of
$M$ and $Z$ for the susy vortex at the one-loop level is a
mystery because one can count how many fermionic zero modes there
are in this model and it has been claimed that there are two,
rather than (as expected) one. With two fermionic zero modes the
equality $M=Z$ could indeed not be explained; it could be due to
some as yet not known symmetry, or it might just be an accident.
However, we have shown that the second fermionic zero mode is not
normalizable at the origin. Hence there is only one fermionic
zero mode, and this means that the equality $M=Z$ which we
obtained by detailed calculations using quantum field theory for
extended objects, is, in fact, a direct consequence of ordinary
supersymmetry at the quantum level.

So far we discussed ordinary dimensional regularization with
$n\geq n_0$. We already discussed that dimensional reduction with
$n\leq n_0$ preserves susy. The number of field components of
fermions and bosons remains fixed (and one should treat some
vector bosons as $\epsilon$ scalars as we discussed). The identity
$\gamma^\mu\gamma_\rho\gamma_\mu =0$ remains valid because the
index $\mu$ keeps running from $0$ to 2, instead of from 0 to
$n_0$. (If this index would have been due to a derivative such as
$j_\mu=\partial_\mu \varphi\psi$ then $\mu$ should only run up to
$n$, but $j_\mu=\partial \!\!\!/\varphi\gamma_\mu\psi$ and so
$\mu$ must run up to $n_0$). This raised a problem many decades
ago: it seemed that there was no conformal susy anomaly in
dimensional reduction. Some people proposed wild solutions:
breakdown of cyclicity of the trace operation or other drastic
measures. Actually, the solution is conventional, although
subtle: there are {\bf evanescent counter terms} \cite{Bonneau}
for the currents.
These are counter terms such as $\Delta j_\mu
=\frac{1}{\epsilon}\partial \!\!\!/\varphi\gamma_{\hat{\mu}}\psi$
where the index $\hat{\mu}$ only runs over $\epsilon$ values
(namely from $n$ to $n_0$). One cannot write $\gamma_{\hat{\mu}}$
itself as $\epsilon$ times a finite quantity, but inside loop
graphs the effect of $\gamma_{\hat{\mu}}$ is to supply a factor
$\epsilon$. Thus $\Delta j_\mu$ yields finite contributions. If
one requires that ordinary susy is preserved, one must satisfy
$\partial^\mu j_\mu ({\rm susy})=0$ in dimensional reduction, and
then  one must renormalize the susy current by adding an
evanescent counter term. {\bf This counter term, and not the
original loop graph, yields the anomaly}. In ordinary dimensional
regularization the situation is just the reverse: there the loop
graph yields the anomaly (as discussed in text books) and the
counter terms do not contribute to the anomaly. One can construct
the whole anomaly multiplet, and one finds then that the
evanescent counter term $\Delta j_\mu$ in the conformal susy
current yields a finite nonsingular contribution to the central
charge anomaly. This finite term is the anomaly in $Z$, and  the
value of this anomaly is the same as that obtained from ordinary
dimensional regularization. To avoid misunderstanding: we also directly
computed this anomaly using dimensional reduction, but as the
preceding discussion shows, one can also obtain it by making susy
transformations of the anomaly $\gamma\cdot j ({\rm susy})$.

We have written several papers on these subjects, and also
published some 
reviews \cite{Goldhaber:2002sh,Goldhaber:2004kn}. 
For a gentle introduction we
recommend \cite{Goldhaber:2002sh}. 
In the remaining sections we focus on the kink,
vortex, and monopole, respectively,
using susy-preserving dimensional regularization methods.

\section{The (susy) kink.}

The calculation of quantum corrections to the mass of
a supersymmetric (susy) kink 
and to its central charge 
has proved to be a surprisingly subtle problem, and it
took protracted struggles to fully understand it in the various
methods that had been employed.

Initially it was thought that supersymmetry would
lead to a complete cancellation of quantum corrections \cite{D'Adda}
and thereby guarantee Bogomolnyi-Prasad-Sommerfield
(BPS) saturation at the quantum level.
Then, by considering a kink-antikink system in a finite
box and regularizing the ultraviolet divergences by
a cutoff in the number of the discretized modes, 
Schonfeld \cite{Schonfeld:1979hg}
found that there is a nonzero, negative
quantum correction at one-loop level, $\Delta M^{(1)}=-m/(2\pi)$.
Most of the subsequent literature
\cite{KaulChatterjeeYamagishi} 
considered instead a single kink directly, using (usually implicitly)
an energy-momentum cutoff which gave
again a null result. A direct calculation of the
central charge \cite{Imbimbo:1984nq} also gave a null result, apparently
confirming a conjecture
of Witten and Olive \cite{Witten:1978mh}
that BPS saturation in the minimally susy 1+1 dimensional case
would hold although arguments on multiplet shortening naively do not
seem to apply.

In Ref.~\cite{Rebhan:1997iv} two of the present authors
noticed a surprising dependence on the regularization method,
even after the renormalization conditions have been fully fixed.
In particular it was found that the
naive energy-momentum cutoff
as used in the susy case spoils the integrability of
the bosonic sine-Gordon model \cite{Dashen:1975hd}. 
Using a mode regularization
scheme and periodic boundary conditions in a finite box
instead led to a susy kink mass correction
$\Delta M^{(1)}=+m(1/4-1/2\pi)>0$ (obtained previously also in
Ref.~\cite{Uchiyama:1986gf})
which together with
the null result for the central charge appeared to
be consistent with the BPS bound, but implying nonsaturation.
Subsequently it was found by two of us together with
Nastase and Stephanov \cite{Nastase:1998sy} that the
traditionally used periodic boundary conditions are questionable. Using
instead topological boundary conditions which are invisible
in the topological and in the trivial sector together with 
a ``derivative regularization''\footnote{In 
mode regularization it turns out that one has to average over
sets of boundary conditions to cancel both localized
boundary energy and delocalized momentum 
\cite{Goldhaber:2000ab,Goldhaber:2002mx}.}
indeed led to a different result,
namely that originally obtained by Schonfeld \cite{Schonfeld:1979hg},
which however appeared to be in conflict with the BPS inequality
for a central charge without quantum corrections.

Since this appeared to be a pure one-loop effect,
Ref.~\cite{Nastase:1998sy} proposed the
conjecture that it may be formulated in terms of a
topological quantum anomaly.
It was
then shown by Shifman et al. \cite{Shifman:1998zy}, using a susy-preserving
higher-derivative regularization method, 
that there is indeed an anomalous contribution to the
central charge balancing the quantum corrections to the
mass so that BPS saturation remains intact. In fact, it was later
understood that multiplet shortening does in fact occur even
in minimally susy 1+1 dimensional theories, giving rise
to single-state supermultiplets \cite{Losev}.

Both results, the nonvanishing mass correction and thus the necessity
of a nonvanishing correction to the central charge,
have been confirmed by a number of different methods 
\cite{Graham:1998qq,Litvintsev:2000is,Goldhaber:2000ab,Goldhaber:2001rp,Wimmer:2001yn,Bordag:2002dg,Goldhaber:2002mx}
validating also the finite mass formula in terms of only
the discrete modes derived in 
Refs.~\cite{Casahorran:1989vd,Boya:1990fp} based on the
method of \cite{Cahill:1976im}.
However, some authors claimed a nontrivial quantum correction
to the central charge \cite{Graham:1998qq,Casahorran:1989nr}
apparently without the need of the anomalous term
proposed in Ref.~\cite{Shifman:1998zy}.

In \cite{Rebhan:2002uk}, we have shown
that a particularly simple and elegant regularization scheme
that yields the correct quantum mass of the susy kink
is dimensional regularization, if the kink is embedded
in higher dimensions as a domain wall
\cite{Bollini}.
Such a scheme was not considered before for the susy kink
because both susy and the existence of finite-energy
solutions seemed to tie one to one spatial dimension.


In \cite{Rebhan:2002yw} we then showed 
the 2+1 dimensional domain wall is BPS saturated through a
nontrivial quantum correction to the momentum in the extra
dimension.
This nontrivial correction is made
possible by the fact that the 2+1 dimensional theory spontaneously breaks
parity, which also allows the appearance of domain wall fermions
of only one chirality.
By dimensionally reducing to 
1+1 dimensions, this parity-violating contribution
to the extra momentum turns out to provide an
anomalous contribution to the central charge as postulated
in Ref.~\cite{Shifman:1998zy}, thereby giving a novel physical
explanation of the latter. 
This is in line with the well-known fact that central charges of
susy theories can be reinterpreted as momenta in higher dimensions.

Hence, in the case of the susy kink, dimensional regularization is seen
to be compatible with susy invariance only at the expense of a 
spontaneous parity
violation, which in turn allows nonvanishing quantum corrections
to the extra momentum in one higher spatial dimension.
On the other hand, the surface term that usually 
exclusively provides the central charge does not receive quantum 
corrections in dimensional regularization, by the same reason
that led to null results previously in other schemes
\cite{Imbimbo:1984nq,Rebhan:1997iv,Nastase:1998sy}.
The nontrivial anomalous quantum correction to the central charge operator
is thus seen to be entirely the remnant of the spontaneous parity violation
in the higher-dimensional theory in which a susy kink
can be embedded by preserving minimal susy.



\subsection{The model}

The real $\ph^4$ model in 1+1 dimensions with spontaneously
broken $Z_2$ symmetry ($\ph\to-\ph$) has topologically
nontrivial finite-energy solutions called ``kinks'' which
interpolate between the two degenerate vacuum states $\ph=\pm v$.
It has a minimally supersymmetric extension 
\cite{DiVecchia:1977bs}
\bel{Lss}
\CL=-\2\left[ (\6_\m\ph)^2+U(\ph)^2+\5\psi\g^\m\6_\m\psi+
U'(\ph)\5\psi\psi \right]
\ee
where $\psi$ is a Majorana spinor, $\5\psi=\psi^{\mathrm T} C$ with
$C\gamma^\mu=-(\gamma^\mu)^TC$. We shall use
a Majorana representation of the Dirac matrices with $\g^0=-i\t^2$,
$\g^1=\t^3$, and $C=\t^2$ in terms of the
standard Pauli matrices $\tau^k$ so that $\psi={\psi^+\choose\psi^-}$ with
real $\psi^+(x,t)$ and $\psi^-(x,t)$.

The $\ph^4$ model 
is defined as the special case
\be\label{Uphi}
U(\ph)=\sqrt{\l\02}\(\ph^2-v_0^2\),\qquad v_0^2\equiv \mu_0^2/\l
\ee
where the $Z_2$ symmetry of the susy action also involves the fermions
according to $\ph\to-\ph, \psi\to\gamma^5\psi$ with 
$\gamma^5=\gamma^0\gamma^1$.
A classical kink at rest at $x=0$ which interpolates
between the two vacua $\ph=\pm v_0$ is given by \cite{Raj:Sol}
\bel{Ksol}
\ph_{K}=v_0 
\tanh\(\mu_0x/\sqrt2\).
\ee

At the quantum level we have to renormalize, and we shall employ
the simplest possible scheme\footnote{See \cite{Rebhan:2002uk} for
a detailed discussion of more general renormalization schemes
in this context.} which consists of putting all
renormalization constants to unity except for a mass counterterm
chosen such that tadpole diagrams cancel completely in
the trivial vacuum. At the one-loop
level and using dimensional regularization this gives
\bea\label{deltamu2}
\delta \mu^2 &=& \lambda\, \delta v^2 = \lambda
\int{dk_0d^{d}k\0(2\pi)^{d+1}}{-i\0k^2+m^2-i\epsilon}\nn
&=& \lambda
\int{d^{d}k\0(2\pi)^d}{1\02[\vec k^2+m^2]^{1/2}},
\eea
where $m=U'(v)=\sqrt{2}\mu$ is the mass of elementary bosons and
fermions and $k^2=\vec k^2-k_0^2$.

The susy invariance of the model 
under $\delta\varphi=\bar\epsilon\psi$ and $\delta\psi=
(\not\!\partial\ph-U)\epsilon$
(with $\mu_0^2$ replaced by
$\mu^2+\delta\mu^2$) leads to the on-shell conserved Noether
current
\be\label{susyj}
j^\mu=-(\not\!\partial\ph+U(\ph))\gamma^\mu\psi
\ee
and two conserved charges $Q^\pm=\int dx\,j^{0\pm}$.

The model (\ref{Lss}) is equally supersymmetric in 2+1 dimensions,
where we use $\gamma^2=\tau^1$. The same renormalization scheme
can be used,
only the renormalization constant (\ref{deltamu2}) has
to be evaluated for $d=2-\epsilon$ in place of $d=1-\epsilon$
spatial dimensions.

While classical kinks in 1+1 dimensions 
have finite energy (rest mass) $M=m^3/\lambda$,
in (noncompact) 2+1 dimensions there exist no longer solitons of
finite-energy. Instead one can have (one-dimensional) domain walls with
a profile given by (\ref{Ksol}) which have finite surface (string) tension
$M/L=m^3/\lambda$. With
a compact extra dimension one can of course use these configurations
to form ``domain strings'' of finite total energy proportional
to the length $L$ of the string when
wrapped around the extra dimension. 

The 2+1 dimensional case is different also with respect to the
discrete symmetries of (\ref{Lss}). In 2+1 dimensions, 
$\gamma^5=\gamma^0\gamma^1\gamma^2 = \pm {\bf 1}$ corresponding
to the two inequivalent choices available for $\gamma^2=\pm\tau^1$
(in odd space-time dimensions the Clifford algebra has
two inequivalent irreducible representations).
Therefore, the sign of the fermion mass (Yukawa) term can no longer
be reversed by $\psi\to\gamma^5\psi$ and there is no longer
the $Z_2$ symmetry  $\ph\to-\ph, \psi\to\gamma^5\psi$.

What the 2+1 dimensional model does break spontaneously is instead
{\em parity}, which corresponds
to changing the sign of one of the spatial coordinates.
The Lagrangian is invariant under $x^m \to -x^m$ for
a given spatial index $m=1,2$ together with $\ph\to-\ph$ (which
thus is a pseudoscalar) and $\psi\to\gamma^m \psi$.
Each of the trivial vacua breaks these invariances spontaneously,
whereas a kink background in the $x^1$-direction with
$\ph_K(-x^1)=-\ph_K(x^1)$ is symmetric with respect to
$x^1$-reflections, but breaks $x^2=y$ reflection invariance.
 

\subsection{Susy algebra}

The susy algebra for the 1+1 and the 2+1 dimensional cases can
both be covered by starting from 2+1 dimensions, the 1+1 dimensional
case following from reduction by one spatial dimension.


In 2+1 dimensions one
obtains classically \cite{Gibbons:1999np}
\begin{eqnarray}
  \label{eq:3dsusy}
&& \{Q^{\alpha},\bar{Q}_{\beta}\}=2i(\gamma^M)^{\alpha}{}_{\beta}P_M\ ,
          \quad (M=0,1,2)\nonumber\\
    &&= 2i(\gamma^0H+\gamma^1(\tilde{P}_x+\tilde{Z_y})
    +\gamma^2(\tilde{P}_y-\tilde{Z}_x))^\alpha{}_\beta,
\end{eqnarray}%
where we separated off two surface terms $\tilde Z_m$ in defining
\begin{eqnarray}
  \label{eq:Ptilde}
\tilde P_m = \int d^dx  \tilde{\mathcal{P}}_m, \;
  &&\tilde{\mathcal{P}}_m=\dot\ph\,\partial_m\ph
      -\2(\bar{\psi}\gamma^0\partial_m\psi),\\
  \label{eq:Ztilde}
\tilde Z_m = \int  d^dx  \tilde{\mathcal{Z}}_m, \;
  &&\tilde{\mathcal{Z}}_m=U(\ph) \partial_m\ph = \partial_m W(\ph)
\end{eqnarray}
with $W(\ph)\equiv\int d\ph\, U(\ph)$.

Having a kink profile in the $x$-direction, which satisfies the
Bogomolnyi equation $\partial_x \ph_K=-U(\ph_K)$, one finds
that with our choice of Dirac matrices
\bea
  \label{eq:qpm}
  &&Q^{\pm}=\int d^2x[(\dot\ph\mp\partial_y\ph)\psi^\pm
      +(\partial_x\ph\pm U(\ph))\psi^\mp],\quad\\
&&\{Q^\pm,Q^\pm\}=2(H \pm (\tilde Z_x - \tilde P_y)),
\eea
and the charge $Q^+$ 
leaves the topological (domain-wall) vacuum $\ph=\ph_K$, $\psi=0$
invariant.
This corresponds to classical
BPS saturation, since with $P_x=0$ and $\tilde P_y=0$
one has $\{Q^+,Q^+\}=2(H+\tilde Z_x)$ and, indeed, with a kink domain wall
$\tilde Z_x/L^{d-1}=W(+v)-W(-v)=-M/L^{d-1}$.

At the quantum level, hermiticity of $Q^\pm$ implies
\be\label{HPyineq}
\< s|H|s\> \ge |\< s|P_y|s\>|  \equiv |\langle s|(\tilde P_y-\tilde Z_x)
|s\rangle|.
\ee
This inequality is saturated when
\be
Q^+\left|s\>=0
\ee
so that BPS states correspond to massless states $P_M P^M=0$
with $P_y=M$ for a kink domain wall in the $x$-direction,
however with infinite momentum and energy unless the $y$-direction
is compact with finite length $L$.


Classically, the susy algebra in 1+1 dimensions is obtained from
(\ref{eq:3dsusy}) simply by dropping $\tilde P_y$
as well as $\tilde Z_y$ so that $P_x\equiv\tilde P_x$.
The term $\gamma^2 \tilde Z_x$ remains, however, with
$\gamma^2$ being the nontrivial $\gamma^5$ of 1+1 dimensions.
The susy algebra simplifies to
\be
\{Q^\pm,Q^\pm\} = 2(H\pm Z),\quad \{Q^+,Q^-\}=2P_x
\ee
and one has the inequality
\be
\<s|H|s\> \ge |\<s|Z|s\>|
\ee
for any quantum state $s$. BPS saturated states have
$Q^+\left|s\>=0$ or $Q^-\left|s\>=0$, corresponding to
kink and antikink, respectively, and preserve half of the
supersymmetry.

\subsection{Fluctuations}

In a kink (or kink domain wall) background one spatial
direction is singled out and we choose this to be along $x$.
The direction orthogonal to the kink direction (parallel
to the domain wall) will be denoted by $y$.

The quantum fields can then be expanded in the
analytically known kink eigenfunctions \cite{Raj:Sol} times plane waves in
the extra dimensions. For the bosonic fluctuations we have
$[-\square+(U{'}^2+UU{''})]\eta=0$ which is solved by
\bea
  \label{eq:bosfluct}
\eta=\int\frac{d^{d-1}\ell}{(2\pi)^{\frac{d-1}{2}}}
\intsum\frac{dk}{\sqrt{4\pi\omega}}\!\!\!\!\!\!
    &&\left(a_{k,\ell}\ e^{-i(\omega t-\ell y)}\phi_k(x)\right.\nn&&
+\left.
      a^{\dagger}_{k,\ell}\ e^{i(\omega t-\ell y)}\phi^{\ast}_k(x)\right).
\qquad\eea
The kink eigenfunctions $\phi_k$ are 
normalized according to $\int dx|\phi|^2=1$ for the discrete states 
and to Dirac distributions for the continuum states
according to $\int dx\, \phi_k^*\phi_{k'}=2\pi\delta(k-k')$.
The mode energies are 
$\omega=\sqrt{\omega_k^2+\ell^2}$
where $\omega_k$ is the energy in the 1+1-dimensional case.   

The canonical equal-time commutation relations 
$[\eta(\vec{x}),\dot{\eta}(\vec{x}')]=i\delta(\vec{x}-\vec{x}')$
are fulfilled with
\begin{equation}
  \label{eq:ETC}
  [a_{k,\ell},a^{\dagger}_{k',\ell'}]=\delta_{kk'}\delta(\ell-\ell'),
\end{equation}%
where for the continuum states $\delta_{k,k'}$ becomes a Dirac delta. 

For the fermionic modes which satisfy the Dirac equation $[\ds+U{'}]\psi=0$
one finds
\begin{eqnarray}
  \label{eq:ferm}
&& \psi=\psi_0+\nn&&\int\frac{d^{d-1}\ell}{(2\pi)^{\frac{d-1}{2}}}
\intsum'\frac{dk}{\sqrt{4\pi\omega}}
 \Bigl[b_{k,\ell}\  e^{-i(\omega t-\ell y)}
      { {\scriptstyle\sqrt{\omega+\ell}}\ \phi_k(x)\choose
                    {\scriptstyle\sqrt{\omega-\ell}}\ is_k(x)} \nn
&&\qquad\qquad\qquad\qquad\qquad\qquad
            + b^{\dagger}_{k,\ell}\ (c.c.)\Bigr],
\eea
with
\be
 \psi_0=\int\frac{d^{d-1}\ell}{(2\pi)^{\frac{d-1}{2}}}b_{0,\ell}\ e^{-i\ell (t- y)}
   {\phi_0\choose 0},\quad b^{\dagger}_0(\ell)=b_0(-\ell).
\ee
Thus, the fermionic zero mode\footnote{By a slight abuse of notation
we shall always label this by a subscript $0$, but this should not be
confused with the threshold mode $k=0$ (which does not appear
explicitly anywhere below).}
of the susy kink turns into
massless modes located on the domain wall, which have only one
chirality, forming a Majorana-Weyl domain wall fermion 
\cite{Rebhan:2002uk,Callan}.\footnote{The 
mode with $\ell=0$
corresponds in 1+1 dimensions to the zero mode of the susy kink.
It has to be counted as half a degree of freedom in mode
regularization \cite{Goldhaber:2000ab}. For dimensional
regularization such subtleties do not play a role because
the zero mode only gives scaleless integrals and these vanish.}

For the massive modes the Dirac equation relates the eigenfunctions
appearing in
the upper and the lower components of the spinors as follows:
\begin{equation}
  \label{eq:sin}
  s_k=\frac{1}{\omega_k}(\partial_x+U{'})\phi_k
     =\frac{1}{\sqrt{\omega^2-\ell^2}}(\partial_x+U{'})\phi_k,
\end{equation}%
so that the function $s_k$ is the 
SUSY-quantum mechanical \cite{Witten:1982df} partner state of 
$\phi_k$ and thus coincides with the eigen modes of the 
sine-Gordon model (hence the notation) \cite{CooKS:QM}.
With (\ref{eq:sin}), their normalization is the same as that of
the $\phi_k$.

The canonical equal-time anti-commutation relations 
$\{\psi^{\alpha}(\vec{x}),\psi^{\beta}(\vec{x}')\}=\delta^{\alpha\beta}
\delta(\vec{x}-\vec{x}')$ 
are 
satisfied if 
\begin{eqnarray}
  \label{eq:fermetc}
  \{b_0(\ell),b^{\dagger}_0(\ell')\}&=&\{b_0(\ell),b_0(-\ell')\}=\delta(\ell-\ell'),
   \nonumber\\
  \{b_{k,\ell},b^{\dagger}_{k',\ell'}\}&=&\delta_{k,k'}\delta(\ell-\ell'),
\end{eqnarray}%
and again the $\delta_{k,k'}$ becomes a Dirac delta for the continuum states. 
The 
algebra (\ref{eq:fermetc}) 
and the solution for the massless mode (\ref{eq:ferm})
show that the operator $b_0(\ell)$  creates right-moving
massless states on the wall when $\ell$ is negative and 
 annihilates them for positive  momentum $\ell$. 
Thus only massless states with 
momentum in  the positive $y$-direction can be created. 
Changing the representation 
of the gamma matrices by $\gamma^2\rightarrow-\gamma^2$, 
which is inequivalent to 
the original one, reverses the situation. 
Now only massless states with momenta in 
the positive $y$-direction exist. Thus depending on the representation
of the Clifford algebra one chirality of the domain wall fermions
is singled out. This is a reflection of the spontaneous violation 
of parity when embedding the susy kink as a domain wall in 2+1 dimensions.

Notice that in (\ref{eq:ferm}) $d$ can be only 2 or 1, for which
$\ell$ has 1 or 0 components, so for strictly $d=1$ $\ell\equiv0$. 
In order to have a susy-preserving dimensional regularization
scheme by dimensional reduction, we shall start from $d=2$ spatial
dimensions, and then make $d$ continuous and smaller than 2.

\subsection{Energy corrections}

Using the mode expansions in the Hamiltonian expanded to second
order in quantum fluctuations, one finds that 
the bosonic and fermionic contributions combine into
\bea\label{zeropoints}
&&\int dx\, d^{d-1}y\, \langle\mathcal{H}^{(2)}\rangle \nn&&=
{L^{d-1}\02}\int dx\int\frac{d^{d-1}\ell}{(2\pi)^{d-1}}
\intsum {dk\02\pi}\frac{\omega}{2}(|\phi_k|^2
   -|s_k|^2).\qquad
\eea
In these expressions, the massless modes (which correspond to the zero mode
of the 1+1 dimensional kink) can be dropped in dimensional regularization
as scaleless and thus vanishing contributions, and the massive discrete
modes cancel between bosons and fermions.\footnote{The zero mode
contributions in fact do not cancel by themselves
between bosons and fermions, because the
latter are chiral. This noncancellation is in fact crucial in energy cutoff
regularization (see Ref.~\cite{Rebhan:2002uk}).}
Carrying out the $x$-integration over the continuous mode functions
gives a difference of spectral densities, namely
\be\label{specdiff}
\int dx (|\phi_k(x)|^2-|s_k(x)|^2)=-\theta'(k)
=-{2m\0k^2+m^2},
\ee
where $\theta(k)$ is the additional phase shift of the mode functions
$s_k$ compared to $\phi_k$.

Combining (\ref{zeropoints}) and (\ref{specdiff}), and adding
in the counterterm contribution from (\ref{deltamu2})
leads to a simple integral
\bea\label{Mkink}
{\Delta M^{(1)}\0L^{d-1}}&=&-{1\04}\int {dk\,d^{d-1}\ell\0(2\pi)^d}
{\omega}\,\theta'(k)+m\delta v^2\nonumber\\
&=&-{1\04}\int {dk\,d^{d-1}\ell\0(2\pi)^d}{\ell^2\0\omega}\theta'(k)\nn
&=&-{2\0d}{\Gamma({3-d\02})\0(4\pi)^{d+1\02}}\, m^d.
\eea
This reproduces the correct known result for the susy kink mass
correction $\Delta M^{(1)}=-m/(2\pi)$ (for $d=1$) and
the surface (string) tension of the
2+1 dimensional susy kink domain wall 
$\Delta M^{(1)}/L=-m^2/(8\pi)$ (for $d=2$) 
\cite{Rebhan:2002uk}.


\subsection{Anomalous contributions to the central charge}
\label{sec33}

In a kink (domain wall) background with only nontrivial $x$ dependence,
the central charge density $\tilde \mathcal Z_x$ receives
nontrivial contributions.
Expanding $\tilde \mathcal Z_x$ 
around the kink background gives
\bea
  \label{eq:Znaiv}
  \tilde{\mathcal{Z}}_x&=&U\partial_x\ph_K-\frac{\delta\mu^2}{\sqrt{2\lambda}}
        \partial_x\ph_K+\partial_x(U\eta)+\2\partial_x(U{'}\eta^2)\nn&&
        +O(\eta^3). 
\eea
Again only the part quadratic in the fluctuations contributes to
the integrated quantity at one-loop order\footnote{Again, this does not hold
for the central charge density locally 
\cite{Shifman:1998zy,Goldhaber:2001rp}.}.
However, using $U{'}(x=\pm\infty)=\pm m$ this leads just to the contribution 
\begin{eqnarray}
  \label{eq:surface}
&&  \int dx\langle\2\partial_x(U{'}\eta^2)\rangle=
    \2 U{'}\langle\eta^2\rangle|_{-\infty}^{\infty}\nn&&=
    m\int\frac{d^{d-1}\ell}{(2\pi)^{d-1}}\int\frac{dk}{2\pi}\frac{1}{2\omega}
\equiv m \delta v^2,
\end{eqnarray}%
which matches
precisely the counterterm $\propto \delta\mu^2$ 
from requiring vanishing tadpoles.
Straightforward application of the rules of
dimensional regularization thus leads to a null result for
the net one-loop correction to $\langle \tilde Z_x\rangle $ in the same way
as found in Refs.~\cite{Imbimbo:1984nq,Rebhan:1997iv,Nastase:1998sy}
in other schemes.

On the other hand, by considering the less singular
combination $\langle H+\tilde Z_x\rangle $ and showing that it vanishes exactly, 
it was concluded in Ref.~\cite{Graham:1998qq}
that $\langle \tilde Z_x\rangle $ has to compensate any nontrivial result
for $\langle H\rangle $, which in Ref.~\cite{Graham:1998qq} was obtained
by subtracting successive Born approximations for scattering
phase shifts. In fact, Ref.~\cite{Graham:1998qq} explicitly
demonstrates how to rewrite $\langle \tilde Z_x\rangle $ into $-\langle H\rangle $,
apparently without the need for the anomalous terms in the quantum central
charge operator
derived in Ref.~\cite{Shifman:1998zy}.

The resolution of this discrepancy is that Ref.~\cite{Graham:1998qq}
did not regularize $\langle \tilde Z_x\rangle $ and the manipulations
needed to rewrite it as $-\langle H\rangle $ (which eventually is
regularized and renormalized) are ill-defined.
Using dimensional regularization one in fact obtains
a nonzero result for $\langle H+\tilde Z_x\rangle $, apparently
in violation of susy.

However, dimensional regularization by embedding the kink
as a domain wall in (up to) one higher dimension, which
preserves susy, instead leads to 
\be
\<H+\tilde Z_x-\tilde P_y\>=0,
\ee
i.e. the saturation of (\ref{HPyineq}), as we shall now verify.

The bosonic contribution to $\langle \tilde P_y\rangle $ involves
\begin{equation}
  \label{eq:pybos}
  \2\langle\dot{\eta}\partial_y\eta+\partial_y\eta\dot{\eta}\rangle=
  -\int\frac{d^{d-1}\ell}{(2\pi)^{d-1}}\intsum {dk\02\pi}\ 
\frac{\ell}{2}|\phi_k(x)|^2.
\end{equation}%
The $\ell$-integral factorizes and gives zero both because it is
a scale-less integral and because the integrand is odd in $\ell$.
Only the fermions turn out to give interesting contributions:
\begin{eqnarray}
  \label{eq:py}
  \langle\tilde{\mathcal{P}}_y\rangle&=&
  \frac{i}{2}\langle\psi^{\dagger}\partial_y\psi\rangle\nonumber\\
&=& \2 \int\frac{d^{d-1}\ell}{(2\pi)^{d-1}}\intsum {dk\02\pi}{\ell\02\omega}
 \bigl[(\omega+\ell)|\phi_k|^2\nn&&
\qquad\qquad\qquad\qquad\qquad+(\omega-\ell)|s_k|^2\bigr]\nonumber\\
  &=&\2\int\frac{d^{d-1}\ell}{(2\pi)^{d-1}}\ \ell\ \theta(-\ell)\ |\phi_0|^2+
              \nonumber\\
  &&+\2\int\frac{d^{d-1}\ell}{(2\pi)^{d-1}}\intsum' {dk\02\pi}
   \Bigl( \frac{\ell}{2}(|\phi_k|^2+|s_k|^2)\nn&&
\qquad\qquad\qquad\qquad\quad
    +\frac{\ell^2}{2\omega}(|\phi_k|^2-|s_k|^2)
\Bigr).\qquad
\end{eqnarray}
{}From the last sum-integral we have separated off the contribution
of the zero mode of the kink, which turns into chiral domain wall fermions
for $d>1$. The contribution of the latter no longer vanishes by
symmetry, but the $\ell$-integral is still scale-less and therefore
put to zero in dimensional regularization. The first sum-integral
on the right-hand side is again zero by both symmetry and scalelessness,
but the final term is not. The $\ell$-integration no longer
factorizes because $\omega=\sqrt{k^2+\ell^2+m^2}$.
Integrating over $x$ and using (\ref{specdiff}) one in fact obtains exactly
the same expression as in the one-loop result for the
energy, Eq.~(\ref{Mkink}).

So for all $d\le2$ we have BPS saturation, $\langle H\rangle =|\langle \tilde Z_x-\tilde P_y\rangle |$,
which in the limit $d\to1$, the susy kink, is made possible by 
a nonvanishing $\langle \tilde P_y\rangle $. The anomaly in the central charge
is seen to arise from a parity-violating contribution in $d=1+\epsilon$
dimensions which is the price to be paid for preserving supersymmetry
when going up in dimensions to embed the susy kink as a domain wall.

It is perhaps worth emphasizing that the above results do not
depend on the details of the spectral densities associated with
the mode functions $\phi_k$ and $s_k$. In the integrated quantities
$\langle H\rangle $ and $\langle \tilde P_y\rangle $ only the difference of the
spectral densities as given by (\ref{specdiff}) is responsible for
the nonvanishing contribution. The function $\theta(k)$ therein is
entirely fixed by the form of the Dirac equation in the asymptotic
regions $x\to\pm\infty$ far away from the kink \cite{Rebhan:1997iv}.

\subsection{Dimensional reduction and evanescent counterterms}

We now describe how the central charge anomaly can be recovered
from Siegel's version of dimensional regularization 
\cite{Siegel}
where $n$ is smaller than the dimension of spacetime and where one keeps
the number of field components fixed, but lowers the number of
coordinates and momenta from 2 to $n<2$. At the one-loop level one
encounters 2-dimensional $\delta_\mu^\nu$ coming from
Dirac matrices, and $n$-dimensional $\hat\delta_\mu^\nu$ from
loop momenta. An important concept which is going to play a role are
the evanescent counter\-terms \cite{Bonneau}
involving the factor ${1\0\epsilon}\hat{\hat{\delta}}{}_\mu^\nu
\gamma_\nu\psi$, where $\hat{\hat\delta}{}_\mu^\nu\equiv \delta_\mu^\nu-
\hat\delta_\mu^\nu$ has only $\epsilon=2-n$ nonvanishing components.


In the trivial vacuum, expanding the supercurrent 
$j_\mu=-(\not\!\partial\ph+U(\ph))\gamma_\mu\psi$
into quantum fields yields
\be\label{jmuexp}
j_\mu=-\left(\not\!\partial\eta+U'(v)\,\eta
+
\2 U''(v)\,\eta^2
\right)\gamma_\mu
\psi + {1\0\sqrt{2\lambda}}\delta\mu^2\gamma_\mu\psi.
\ee
Only matrix elements with one external fermion are divergent.
The term involving $U''(v)\eta^2$ in (\ref{jmuexp}) gives rise to
a divergent scalar tadpole that is cancelled 
completely by the counterterm $\delta\mu^2$ (which
itself is due to an $\eta$ and a $\psi$ loop). The only other
divergent diagram is due to the term involving $\not\!\partial\eta$
in (\ref{jmuexp}) and has the form a $\psi$-selfenergy. Its
singular part reads
\bea
\!\!\!\!&&\<0| j_\mu |p\>^{\rm div} \nn&&\!\!\!\! = i U''(v)
\int_0^1 dx \int {d^n \kappa\0(2\pi)^n} {\not\!\kappa \gamma_\mu\! \not\!\kappa
u(p) \over [\kappa^2 + p^2 x(1-x) + m^2]^2}.\qquad
\eea
Using $\hat\delta_\mu^\nu\equiv \delta_\mu^\nu-\hat{\hat\delta}{}_\mu^\nu$
we find that under the integral
$$\not\!\kappa \gamma_\mu\!\! \not\!\kappa = - \kappa^2(\delta_\mu^\lambda-
{2\0n}\hat \delta_\mu^\lambda)\gamma_\lambda=
{\epsilon\0n}\kappa^2\gamma_\mu - {2\0n}\kappa^2
\hat{\hat\delta}{}_\mu^\lambda \gamma_\lambda$$
so that
\be
\<0| j_\mu |p\>^{\rm div} = {U''(v)\02\pi}{\hat{\hat\delta}{}_\mu^\lambda
\0\epsilon} \gamma_\lambda u(p).
\ee
Hence, the regularized one-loop contribution to the susy 
current contains the evanescent operator
\be
j_\mu ^{\rm div} = {U''(\ph)\02\pi}{\hat{\hat\delta}{}_\mu^\lambda
\0\epsilon} \gamma_\lambda  \psi.
\ee
This is by itself a conserved quantity, because all fields
depend only on the $n$-dimensional coordinates, but it
has a nonvanishing contraction with $\gamma^\mu$.
The latter 
gives rise to an anomalous contribution to the 
renormalized conformal-susy current
$\not \!x j_\mu^{\rm ren.}$ where $j_\mu^{\rm ren.}=j_\mu-j_\mu^{div}$,
\be
\partial^\mu (\not\!x j_\mu^{\rm ren.})_{\rm anom.}=-
\gamma^\mu j_\mu^{\rm div}= -{U''\02\pi}\psi.
\ee
(There are also nonvanishing
nonanomalous contributions
to $\partial^\mu (\not\!x j_\mu)$ because our model is not
conformal-susy invariant at the classical level
\cite{Fujikawa}.)

Ordinary susy on the other hand is unbroken; there is no anomaly
in the divergence of $j_\mu^{\rm ren.}$. A susy variation of
$j_\mu$ involves the energy-momentum tensor and the
topological central-charge current $\zeta_\mu$
according to
\be
\delta j_\mu = -2T_\mu{}^\nu \gamma_\nu \epsilon - 2 \zeta_\mu \gamma^5 
\epsilon,
\ee
where classically $\zeta_\mu=\epsilon_{\mu\nu}U
\partial^\nu \varphi$.

At the quantum level,
the counter-term $j_\mu^{\rm ct}=-j_\mu^{\rm div.}$ induces
an additional contribution to the central charge current
\be
\zeta_\mu^{\rm anom}={1\04\pi}
{\hat{\hat\delta}{}_\mu^\nu\0\epsilon}
\epsilon_{\nu\rho}\partial^\rho U'
\ee
which despite appearances is a {\em finite} quantity: using that total
antisymmetrization of the three lower indices has to vanish
in two dimensions gives
\be
\hat{\hat\delta}{}_\mu^\nu
\epsilon_{\nu\rho} = \epsilon \epsilon_{\mu\rho}+
\hat{\hat\delta}{}_\rho^\nu
\epsilon_{\nu\mu} 
\ee
and together with the fact the $U'$ only depends on $n$-dimensional
coordinates this finally yields
\be\label{zetaanom}
\zeta_\mu^{\rm anom}={1\04\pi} \epsilon_{\mu\rho} \partial^\rho U'
\ee
in agreement with the anomaly in the central charge as obtained
previously.

\section{The (susy) vortex.}

We next considered \cite{Rebhan:2003bu} the Abrikosov-Nielsen-Olesen
\cite{Abrikosov:1957sx,Nielsen:1973cs,deVega:1976mi,Taubes:1980tm}
vortex solution of the abelian Higgs model in 2+1 dimensions
which has a supersymmetric extension
\cite{Schmidt:1992cu,Edelstein:1994bb} {\rm (see also
\cite{Gorsky:1999hk,Vainshtein:2000hu})} such that classically
the Bogomolnyi bound \cite{Bogomolny:1976de} is saturated. We
employed our variant of dimensional regularization to the $N=2$
vortex by dimensionally  reducing the  $N=1$ abelian Higgs model
in $3+1$ dimensions. We confirmed the results of
\cite{Schmidt:1992cu,Lee:1995pm,Vassilevich:2003xk} that in a
particular gauge (background-covariant Feynman-'t Hooft) the sums
over zero-point energies of fluctuations in the vortex background
cancel completely, but contrary to
\cite{Schmidt:1992cu,Lee:1995pm} we found a nonvanishing quantum
correction to the vortex mass coming from a finite
renormalization of the expectation value of the Higgs field in
this gauge \cite{Wimmer:2003,Vassilevich:2003xk}. In contrast to
\cite{Schmidt:1992cu}, where a null result for the quantum
corrections to the central charge was stated, we show that the
central charge receives also a net nonvanishing quantum
correction, namely from a nontrivial phase in the fluctuations of
the Higgs field in the vortex background, which contributes to
the central charge even though the latter is a surface term that
can be evaluated far away from the vortex. The correction to the
central charge exactly matches the correction to the mass of the
vortex.

In Ref.~\cite{Lee:1995pm}, it was claimed that the usual
multiplet shortening arguments in favor of BPS saturation would
not be applicable to the $N=2$ vortex since in the vortex
background there would be two rather than one fermionic zero
modes \cite{Lee:1992yc}, leading to two short multiplets which
have the same number of states as one long
multiplet.\footnote{Incidentally, Refs.\
\cite{Lee:1995pm,Lee:1992yc} considered supersymmetric
Maxwell-Chern-Simons theory, which contains the supersymmetric
abelian Higgs model as a special case.} We showed however that the
extra zero mode postulated in \cite{Lee:1995pm} has to be
discarded because its gaugino component is singular, and that
only after doing so there is agreement with the results from
index theorems \cite{Weinberg:1981eu,Lee:1992yc,Hori:2000kt}. For
this reason, standard multiplet shortening arguments do apply,
explaining the BPS saturation at the quantum level that we observe
in our explicit one-loop calculations.

The $N=2$ \SUSY\ vortex in 2+1 dimensions is the solitonic
(finite-energy) solution of the abelian Higgs model which can be
obtained by dimensional reduction from a 3+1-dimensional $N=1$
model. We shall use the latter for the purpose of dimensional
regularization of the 2+1-dimensional model by \SUSY-preserving
dimensional reduction from 3+1 dimensions (where the vortex has
infinite mass but finite energy-density).

\subsection{The model}

The superspace action for the vortex in terms of 3+1-dimensional
superfields contains an $N=1$ abelian vector multiplet and an
$N=1$ scalar multiplet, coupled as usual, together with a
Fayet-Iliopoulos term but without superpotential, \be
\mathcal L = 
\int d^2 \theta\, W^\alpha W_\alpha + \int d^4\theta\, \bar\Phi\,
e^{eV} \Phi + \kappa \int d^4\theta\, V. \ee In terms of
2-component spinors in 3+1 dimensions, the action
reads\footnote{Our conventions are $\eta^{\mu\nu}=(-1,+1,+1,+1)$,
$\chi^{\alpha}= \epsilon^{\alpha\beta}\chi_\beta$ and
$\5\chi^{\dot\alpha}=
\epsilon^{\dot\alpha\dot\beta}\5\chi_{\dot\beta}$ with
$\epsilon^{\alpha\beta}=\epsilon_{\alpha\beta}=
-\epsilon^{\dot\alpha\dot\beta}=-\epsilon_{\dot\alpha\dot\beta}$
and $\epsilon^{12}=+1$. In particular we have $\5\psi_{\.\alpha}=
(\psi_\alpha)^*$ but $\5\psi^{\.\alpha}=- (\psi^\alpha)^*$.
Furthermore, $\5\sigma^\mu=(-\mathbf 1,\vec\sigma)$ with the
usual representation for the Pauli matrices $\vec\sigma$.}
\bea\label{L4dmodel} \mathcal L &=& -{1\04}F_{\mu\nu}^2
+\5\chi^{\.\alpha} i \5\sigma_{\.\alpha\beta}^\mu \6_\mu
\chi^\beta +{1\02}D^2+ (\kappa-e|\phi|^2)D\nn &&-|D_\mu \phi|^2+
\5\psi^{\.\alpha} i \5\sigma_{\.\alpha\beta}^\mu D_\mu \psi^\beta
+|F|^2\nn&&
 +\sqrt2 e \left[ \phi^* \chi_\alpha \psi^\alpha +\phi
\5\chi_{\.\alpha} \5\psi^{\.\alpha} \right], \eea where
$D_\mu=\6_\mu - ieA_\mu$ when acting on $\phi$ and $\psi$, and
$F_{\mu\nu}=\6_\mu A_\nu-\6_\nu A_\mu$. Elimination of the
auxiliary field $D$ yields the scalar potential $\mathcal V=\2
D^2=\2 e^2(|\phi|^2-v^2)^2$ with $v^2\equiv {\kappa/e}$.

In 3+1 dimensions, this model has a chiral anomaly, and in order
to cancel the chiral U(1) anomaly, additional scalar multiplets
would be needed such that the sum over charges vanishes, $\sum_i
e_i=0$. 

In 2+1 dimensions, dimensional reduction gives an $N=2$ model
involving, in the notation of \cite{Lee:1995pm}, a real scalar
$N=A_3$ and two complex (Dirac) spinors $\psi=(\psi^\alpha)$,
$\chi=(\chi^\alpha)$.

Completing squares in the bosonic part of the classical
Hamiltonian density one finds the Bogomolnyi equations and the
central charge \bea \mathcal H&=&{1\04}F_{kl}^2+|D_k \phi|^2+\2
e^2 (|\phi|^2-v^2)^2 \nn &=& \2 |D_k \phi + i\epsilon_{kl} D_l
\phi|^2 +\2 \left(F_{12}+e(|\phi|^2-v^2)\right)^2 \nn&& +{e\02}
v^2 \epsilon_{kl}F_{kl} - i \6_k (\epsilon_{kl}\phi^* D_l \phi)
\eea where $k,l$ are the spatial indices in 2+1 dimensions. The
classical central charge reads \be Z=\int d^2x \, \epsilon_{kl}
\6_k \left( e v^2 A_l - i \phi^* D_l \phi \right), \ee where
asymptotically $D_l\phi$ tends to zero exponentially fast.
Classically, BPS saturation $E=|Z|=2\pi v^2 n$ holds when the BPS
equations $(D_1 \pm iD_2)\phi \equiv D_\pm \phi=0$ and $F_{12}
\pm e(|\phi|^2-v^2)=0$ are satisfied, where the upper and lower
sign corresponds to vortex and antivortex, respectively. The
vortex solution with winding number $n$ is given by
($A_\pm^{\mathrm V} \equiv A_1^{\mathrm V} \pm i A_2^{\mathrm V}$)
\be\label{vortexsolution} \pv = e^{in\theta} f(r), \quad
eA_+^{\mathrm V} = -i e^{i\theta}{a(r)-n\0r}, 
\ee
where $f'(r)={a\0r}f(r)$ and $a'(r)=r e^2(f(r)^2-v^2)$ with
boundary conditions \cite{Taubes:1980tm} \bea &a(r\to\infty)=0,
&f(r\to\infty)=v,\nn &a(r\to0)=n+O(r^2), &f(r\to0)\,\propto\,
r^n+O(r^{n+2}).\qquad\eea

\subsection{Fluctuation equations}

For the calculation of quantum corrections to a vortex solution we
decompose $\phi$ into a classical background part $\pv$ and a
quantum
part $\eta$. Similarly, $A_\mu$ 
is decomposed as $A_\mu^{\mathrm V}+a_\mu$, where only
$A_\mu^{\mathrm V}$ with $\mu=1,2$ is nonvanishing.
We use a background $R_\xi$ \cite{'tHooft:1971rn
} gauge fixing term which is quadratic in the quantum gauge
fields, \be\label{Lgfix} \mathcal L_{\rm g.fix}= -{1\02\xi}
(\6_\mu a^\mu - ie\xi(\pv \eta^* - \pv^* \eta ))^2. \ee The
corresponding Faddeev-Popov Lagrangian reads \be \mathcal L_{\rm
ghost}= b \left( \6_\mu^2 - e^2 \xi \left\{ 2\,|\pv|^2 + \pv
\eta^* + \pv^* \eta  \right\} \right) c\,. \ee

The fluctuation equations in 2+1 dimensions have been given in
\cite{Schmidt:1992cu,Lee:1995pm} for the
choice $\xi=1$ (Feynman-`t Hooft gauge
) which leads to important simplifications. We shall mostly use
this gauge choice when considering fluctuations in the solitonic
background, but will carry out renormalization in the trivial
vacuum for general $\xi$ to exhibit some of the intermediate
gauge dependences.

Because we are going to consider dimensional regularization by
dimensional reduction from the 3+1 dimensional model, we shall
need the form of the fluctuation equations with derivatives in
the $x^3$ direction included. (This one trivial extra dimension
will eventually be turned into $\epsilon\to0$ dimensions.)

In the `t Hooft-Feynman gauge, the part of the bosonic action
quadratic in the quantum fields reads 
\bea \mathcal L_{\rm
bos}^{(2)}
&=& -\2 (\6_\mu a_\nu)^2 
- e^2 |\pv|^2 a_\mu^2 \nn&&
- |D_\mu^{\mathrm V} \eta|^2 - e^2 (3|\pv|^2-v^2) |\eta|^2\nn&&
-2ie a^\mu \left[\eta^* D_\mu^{\mathrm V} \pv - \eta
(D_\mu^{\mathrm V} \pv)^* \right].\quad \eea In the trivial
vacuum, which corresponds to $\pv\to v$ and $A_\mu^{\mathrm V}
\to 0$, the last term vanishes, but in the solitonic vacuum it
couples the linearized field equations for the fluctuations
$B\equiv(\eta,a_+/\sqrt2)$
with $a_+=a_1+ia_2$ 
to each other according to ($k=1,2$) 
\bea\label{Bfluceq}
&&(\6_3^2-\6_t^2)B\nn
&&= \left(
\begin{array}{cc}
-(D_k^{\mathrm V})^2+e^2(3|\pv|^2-v^2)  &  i\sqrt2 e (D_- \pv) \\
-i\sqrt2 e (D_- \pv)^* & -\6_k^2+2 e^2 |\pv|^2
\end{array}
\right)B.\nn \eea 
The quartet $(a_3,a_0,b,c)$ with $b,c$ the
Faddeev-Popov ghost fields has diagonal field equations at the
linearized level \be (\6_\mu^2 - 2 e^2 |\pv|^2) Q=0, \quad
Q\equiv(a_3,a_0,b,c). \ee

For the fermionic fluctuations, which we group as $U=\left(
\psi^1 \atop \5\chi^{\.1} \right)$, $V=\left( \psi^2 \atop
\5\chi^{\.2} \right)$, the linearized field equations read
\be\label{UVeqs} LU=i(\6_t+\6_3) V, \quad L^\dagger V =
i(\6_t-\6_3) U, \ee 
with 
\be L= \left(\begin{array}{cc}
iD_+^{\mathrm V} & \sqrt2 e \pv \\
-\sqrt2 e \pv^*  & i\6_-
\end{array}\right)\!,
L^\dagger = \left(\begin{array}{cc}
iD_-^{\mathrm V} & -\sqrt2 e \pv \\
\sqrt2 e \pv^*  & i\6_+
\end{array}\right)\!.
\ee

Iteration shows that $U$ satisfies the same second order equations
as the bosonic fluctuations $B$, \bea
&&L^\dagger L U=(\6_3^2-\6_t^2) U, \quad L^\dagger L B=(\6_3^2-\6_t^2) B\quad\\
&&L L^\dagger V=(\6_3^2-\6_t^2) V, \eea with $L^\dagger L$ given
by (\ref{Bfluceq}), whereas $V$ is governed by a diagonal
equation with \be\label{LLd} LL^\dagger=\left(
\begin{array}{cc}
-(D_k^{\mathrm V})^2+e^2|\pv|^2+e^2 v^2  &  0 \\
0 & -\6_k^2+2 e^2 |\pv|^2
\end{array}
\right). \ee (In deriving these fluctuation equations we used the
BPS equations throughout.)

\subsection{Renormalized mass}

At the classical level, the energy and central charge of vortices
are multiples of $2\pi v^2$ with $v^2=\kappa/e$. Renormalization
of tadpoles, even when only by finite amounts, will therefore
contribute directly to the quantum mass and central charge of the
$N=2$ vortex, a fact that has been overlooked in the original
literature \cite{Schmidt:1992cu,Lee:1995pm} on quantum
corrections to the $N=2$ vortex.\footnote{The nontrivial
renormalization of $\kappa/e$ has however been included in
\cite{Wimmer:2003,Vassilevich:2003xk}.}

Adopting a ``minimal'' renormalization scheme where the scalar
wave function renormalization constant $Z_\phi=1$, the
renormalization of $v^2$ is fixed by the requirement of vanishing
tadpoles in the trivial sector of the 2+1 dimensional model. The
calculation can be conveniently performed by using dimensional
regularization of the 3+1 dimensional $N=1$ model. For the
calculation of the tadpoles we decompose $\phi=v+\eta\equiv
v+(\sigma+i\rho)/\sqrt2$, where $\sigma$ is the Higgs field and
$\rho$ the would-be Goldstone boson.
The gauge fixing term (\ref{Lgfix}) avoids mixed $a_\mu$-$\rho$
propagators, but there are mixed $\chi$-$\psi$ propagators, which
can be
diagonalized 
by introducing new spinors $s=(\psi+i\chi)/\sqrt2$ and
$d=(\psi-i\chi)/\sqrt2$ with mass terms $m(s_\alpha s^\alpha -
d_\alpha d^\alpha)+h.c.$, where $m=\sqrt2 e v$.

The part of the interaction Lagrangian which is relevant for
$\sigma$ tadpoles to one-loop order is given by 
\bea \mathcal
L^{\rm int}_{\sigma-{\rm tadpoles}} &=& e(\chi_\alpha
\psi^\alpha+\5\chi_{\.\alpha} \5\psi^{\.\alpha})\,\sigma
-{em\02}(\sigma^2+\rho^2)\,\sigma\nn&&
-em(a_\mu^2+
\xi b\,c- \delta v^2)\,\sigma, 
\eea 
where $b$ and $c$ are the Faddeev-Popov fields.

The one-loop contributions to the $\sigma$ tadpole thus read
\bea\label{sigmatadpoles}
&&\!\!\!\includegraphics[width=0.45\textwidth]{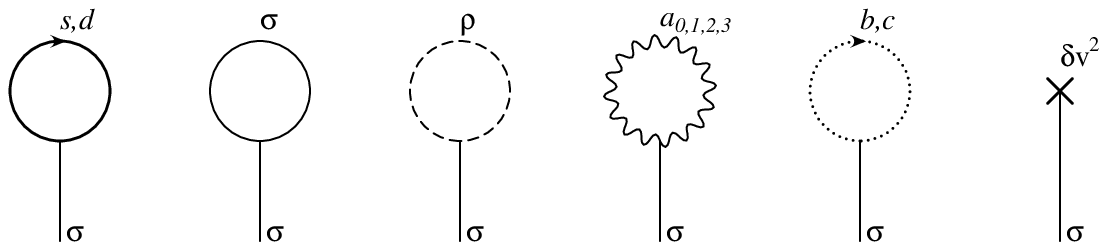} \nn
&& =(-em)\Bigl\{-2 {\rm tr}{\mathbf 1_2}I(m) + \frac{3}{2}I(m) +
\2 I(\xi^{\frac{1}{2}}m) \nn&&
\qquad\qquad\quad + [3I(m)+\xi I(\xi^{\2}m)] - \xi
I(\xi^{\2}m) -\delta v^2 \Bigr\},\quad \nonumber
\eea
where 
\bea I(m)&=&\int
{d^{3+\epsilon}k\0(2\pi)^{3+\epsilon}}{-i\0k^2+m^2}\nn&=&
-{m^{1+\epsilon}\0(4\pi)^{1+\epsilon/2}}
{\Gamma(-\2-{\epsilon\02})\0 \Gamma(-\2)}=-{m\04\pi}+O(\epsilon).\quad
\eea

Requiring that the sum of tadpole diagrams (\ref{sigmatadpoles})
vanishes fixes $\delta v^2$, \be\label{deltav2Rxi} \delta v^2 =
\2\left(I(m)+I(\xi^{1\02}m)\right)\Big|_{D=3} =
-{1+\xi^{1\02}\08\pi}m. \ee Because in dimensional regularization
there are no poles in odd dimensions at the one-loop level, the
result for $\delta v^2$ is finite, but it is nonvanishing.

As it turns out, this is the only contribution to the one-loop
mass correction of the vortex. In the $\xi=1$ gauge
the zero-point energies of the quartet $(a_3,a_0,b,c)$ cancels,
and one is left with
\be\label{zeromodesums}
\2\sum \omega_{\rm bos} - \2\sum \omega_{\rm ferm}
= \sum \omega_U - \sum \omega_V,
\ee

Using dimensional regularization these sums can be made
well defined by replacing all eigen frequencies $\omega_k$ in 2+1
dimensions by $\omega_{k,\ell}=(\omega_k^2+\ell^2)^{1/2}$ where
$\ell$ are the extra momenta. Using index theorems, it has
been shown that the spectral densities for $U$ and $V$ are
equal up to zero modes \cite{Schmidt:1992cu,Lee:1995pm},
and zero modes (massless modes upon embedding)
do not contribute in dimensional regularization.
Hence, $\sum \omega_U - \sum \omega_V=0$, as we have
also verified more directly \cite{Goldhaber:2004kn}, and the only
nonvanishing quantum correction of the vortex mass is from
renormalization. In our ``minimal'' renormalization scheme we thus have
\be\label{MV}
E
= 2\pi |n| (v^2 + \delta v^2|_{\xi=1}) = 
2\pi |n| (v^2 - \frac{m}{4\pi})
.
\ee

\subsection{Central charge}

By starting from the \SUSY\ algebra in 3+1 dimensions
one can derive the central charge in 2+1 dimensions as the
component $T_{03}$ of
\be
T^{\mu\nu}=-{i\04} {\rm Tr} \,\sigma^{\mu \alpha\dot\alpha}\,
\{ \bar Q_{\dot\alpha},J_\alpha^\nu\}
\ee
where $J_\alpha^\nu$ is the \SUSY\ Noether current.

The antisymmetric part of $T^{\mu\nu}$ gives the
standard expression for the central charge density, while
the symmetric part is a genuine momentum in the extra
dimension:
\be
\< Z \> = \int d^2 x \< T^{03} \> 
= \< \tilde Z + \tilde P_3 \>.
\ee
(A similar decomposition is valid for the kink \cite{Rebhan:2002yw}.)

$\tilde Z$ corresponds to the classical expression for the central
charge. Being a surface term, its quantum corrections can be
evaluated at infinity:
\be
\langle \tilde Z \rangle =
\int d^2x \6_k \epsilon_{kl} \langle \tilde\zeta_l \rangle
= \int_0^{2\pi}\!\!\! d\theta \langle \tilde\zeta_\theta \rangle|_{r \to \infty}
\ee
with 
$\tilde\zeta_l=e v^2_0 A_l-i\phi^\dagger D_l \phi$ and $v_0^2=v^2+\delta v^2$.

Expanding in quantum fields $\phi=\pv+\eta$, 
$A=A^{\rm V}+a$ and using that
the classical fields $\pv\to v e^{in\theta}$, $A_\theta^{\rm V}\to n/e$,
$D_\theta^{\rm V} \pv \to 0$ as $r\to \infty$,
we 
obtain to one-loop order 
\bea\label{Zab}
\langle \tilde Z \rangle 
&=& 2\pi n v^2_0 \nn&&- i \int_0^{2\pi}\!\!\! d\theta \left\langle 
(\pv^*+\eta^\dagger) (D_\theta^{\rm V}-ie a_\theta)(\pv+\eta)
\right\rangle|_{r \to \infty} \nn
&=& 2\pi n \{v^2_0 - \langle \eta^\dagger \eta \rangle |_{r \to \infty}\}
\nn&&
- i \int_0^{2\pi}\!\!\! d\theta \bigl\{
\left\langle 
\eta^\dagger \6_\theta \eta 
\right\rangle \nn&&\qquad\qquad
-ie \pv^* \left\langle a_\theta \eta \right\rangle
-ie \pv \left\langle a_\theta \eta^\dagger \right\rangle
\bigr\}|_{r \to \infty}\nn
&\equiv& Z_a+Z_b
\eea
where we have used 
$\langle \eta(r\!\to\!\infty) \rangle \to 0$ (which determines $\delta v^2$),
$\int_0^{2\pi}\! d\theta\langle a_\theta \rangle = 0$, and
$\langle \eta^\dagger \eta a_\theta \rangle = O(\hbar^2)$.

The first contribution, $Z_a$, can be easily evaluated for
arbitrary gauge parameter $\xi$, yielding
\bea
Z_a&=&2\pi n \{v^2_0
-\2(
\langle \sigma\sigma \rangle + \langle \rho\rho \rangle) |_{r \to \infty}\}
\nn&=&
2\pi n \{v^2_0-\2[I(m)+I(\xi^{1\02}m)]\}\nn&=&2\pi n (v_0^2-\delta v^2)=
2\pi n v^2.
\eea
If this was all, the quantum corrections to $Z$
would cancel, just as in the naive
calculation of $Z$ in the \SUSY\ kink \cite{Imbimbo:1984nq,Rebhan:1997iv}.

The second contribution in (\ref{Zab}), however, does not
vanish when taking the limit $r\to\infty$. This contribution
is simplest in the $\xi=1$ gauge, where the $\eta$ and $a_\theta$
fluctuations are governed by the fluctuation equations
(\ref{Bfluceq}). In the limit $r\to\infty$ one has
$|\pv|\to v$ and $D_-\pv\to 0$ exponentially. This
eliminates the contributions from $\left\langle a_\theta \eta \right\rangle$.
However, $D_k^2$, which governs the $\eta$ fluctuations,
contains long-range contributions from the vector
potential. Making a separation of variables in $r$ and $\theta$
one finds that asymptotically 
\be
|D_k^{\mathrm V} \eta|^2 \to |\6_r \eta|^2+{1\0r^2} |(\6_\theta-in)\eta|^2
\ee
so that the $\eta$ fluctuations
have an extra phase factor $e^{in\theta}$ compared to those in the
trivial vacuum. We thus have, in the $\xi=1$ gauge,
\bea
Z_b&=&
- i \int_0^{2\pi}\!\!\! d\theta \bigl\{
\left\langle 
\eta^\dagger \6_\theta \eta 
\right\rangle \nn&&\qquad\qquad
-ie \pv^* \left\langle a_\theta \eta \right\rangle
-ie \pv \left\langle a_\theta \eta^\dagger \right\rangle
\bigr\}|_{r \to \infty}\nn
&=&- i \int_0^{2\pi}\!\!\! d\theta 
\left\langle \eta^\dagger \6_\theta \eta 
\right\rangle_{\xi=1} 
= 2\pi n \left\langle \eta^\dagger \eta \right\rangle_{\xi=1,r \to \infty}
\nn&=& 2\pi n\, \delta v^2\Big|_{\xi=1}\,.
\eea
This is exactly the result for the one-loop correction to the
mass of the vortex in
eq.~(\ref{MV}), implying saturation of the BPS bound 
provided that there are now no anomalous contributions to the
central charge operator 
as there are in the case in the $N=1$ \SUSY\
kink \cite{Rebhan:2002yw}.

In 
dimensional regularization by dimensional
reduction from a higher-dimensional model such anomalous
contributions to the central charge operator come from
a finite remainder of the extra momentum operator \cite{Rebhan:2002yw}. 
The latter is
given by \cite{Lee:1995pm}
\bea
Z_c=\< \tilde P_3\>&\!=\!&
\int d^{2}x \< F_{0i}F_{3i}
+ (D_0 \phi)^\dagger D_3 \phi
+ (D_3 \phi)^\dagger D_0 \phi \right. \nn
&&\qquad\qquad\left.
- i \bar \chi \bar\sigma_0 \6_3 \chi
-i \bar \psi \bar\sigma_0 D_3 \psi \>.
\eea
Inserting mode expansions for the quantum fields one
immediately finds that the bosonic contributions vanish 
because of symmetry in the extra trivial dimension.
However, this is not the case for the fermionic fields,
which have a mode expansion of the form
\bea
\left( U \atop V \right) &=&
\int{d^\epsilon \ell \0 (2\pi)^{\epsilon/2}}
{\Sigma}\!\!\!\!\!\!\int_k
{1\0\sqrt{2\omega}} \Bigl\{
b_{k,\ell}\, e^{-i(\omega t-\ell z)}
\left(  {\sqrt{\omega-\ell}\, u_1 \atop \sqrt{\omega-\ell}\, u_2} 
\atop   {\sqrt{\omega+\ell}\,v_1 \atop \sqrt{\omega+\ell}\,v_2} \right)
\nn&&\qquad\qquad\qquad\qquad
+d_{k,\ell}^\dagger \times(c.c.)
\Bigr\},
\eea
where we have not written out explicitly the zero-modes
(for which $\omega^2=\ell^2$).
The fermionic contribution to $Z_c$ reads, schematically,
\bea
&&Z_c=\< \tilde P_3\>=\nn
&& \int{d^\epsilon \ell \0 (2\pi)^\epsilon}
{\Sigma}\!\!\!\!\!\!\int_k {\ell^2\02\omega}
\int d^2x \left[ |u_1|^2+|u_2|^2-|v_1|^2-|v_2|^2 \right]
\qquad\eea
where $\omega=\sqrt{\omega_k+\ell^2}$, so that the $\ell$ integral
does give a nonvanishing result. However, the $x$-integration
over the mode functions $u_{1,2}(k;x)$ and $v_{1,2}(k;x)$
produces their spectral densities, which cancel up to
zero-mode contributions as we have
seen above\footnote{An explicit calculation which
confirms these cancellations can be found in
\cite{Goldhaber:2004kn}.}, and zero-mode contributions
only produce scaleless integrals
which vanish in dimensional regularization. Hence, $Z_c=0$
and $|Z|=|Z_a+Z_b|=E$, so that the BPS bound is saturated at
the (one-loop) quantum level.

\section{The (susy) monopole.}

We now consider the $N=2$ monopole in 3+1 dimensions,
which has been used by many authors in studies of duality. The
monopole model has more unbroken susy generators than the susy
kink or the vortex, so one runs the risk (or the blessing) of
vanishing quantum corrections. This model has been studied before
in Refs.~\cite{D'Adda:1978mu,Kaul:1984bp,Imbimbo:1985mt} and
while the initial result of vanishing corrections of
Ref.~\cite{D'Adda:1978mu} turned out to be an oversimplification,
Refs.~\cite{Kaul:1984bp,Imbimbo:1985mt} nevertheless arrived at
the conclusion of vanishing quantum corrections, at least in the
simplest renormalization schemes.  
Using susy-preserving dimensional regularization by dimensional
reduction, we have instead recently shown \cite{Rebhan:2004vn} 
that there are nonvanishing but equal quantum
corrections to both mass and central charge of the $N=2$ monopole.
Thus BPS saturation is preserved as required by
multiplet shortening arguments \cite{Witten:1978mh}, but in a nontrivial
manner.


The $N=2$ super-Yang-Mills theory in 3+1 dimensions can be
obtained by dimensional reduction from the (5+1)-dimensional
$N=1$ theory
\cite{Brink:1977bc}
\begin{equation}
  \label{eq:L6d}
  \mathcal L
=- \frac{1}{4} F_{AB}^2 - \bar{\lambda}\Gamma^AD_A\lambda,
\end{equation}
where the indices $A,B$ take the values $0,1,2,3,5,6$ and which
is invariant under \be \delta A_B^a=\bar\lambda^a \Gamma_B
\eta-\bar\eta \Gamma_B \lambda^a ,\quad \delta \lambda^a={1\02}
F^a_{BC}\Gamma^B \Gamma^C \eta.\ee The complex spinor $\lambda$
is in the adjoint representation of the gauge group which we
assume to be SU(2) in the following and
$(D_A\lambda)^a=(\partial_A\lambda+gA_A\times \lambda)^a\break
=\partial_A\lambda^a+g\epsilon^{abc}A_A^b\lambda^c$. Furthermore,
$\lambda$ and $\eta$ satisfy the Weyl condition:
\begin{equation}
  \label{eq:wcon}
  (1-\Gamma_7)\lambda=0\quad \textrm{with} \quad
  \Gamma_7=\Gamma_0\Gamma_1\Gamma_2\Gamma_3\Gamma_5\Gamma_6.
\end{equation}

To carry out the dimensional reduction we write
$A_B=(A_\mu,P,S)$ and choose the following representation of
gamma matrices
\begin{eqnarray}
  \label{eq:6dgam}
  \Gamma_\mu=\gamma_\mu\otimes\sigma_1\ &,&\ \mu=0,1,2,3,\nonumber\\
  \Gamma_5=\gamma_5\otimes\sigma_1\ &,&\ \Gamma_6=\unit\otimes\sigma_2.
\end{eqnarray}
In this representation the Weyl condition (\ref{eq:wcon}) becomes
$\lambda={\psi\choose 0}$, with a complex four-component spinor
$\psi$.\footnote{We use the metric with signature $(-,+,+,+,+,+)$
and $\bar\lambda^a=(\lambda^a)^\dagger i \Gamma^0$, hence
$\bar\psi^a=(\psi^a)^\dagger i \gamma^0$. One can rewrite this
model in terms of two symplectic Majorana spinors in order to
exhibit the $R$ symmetry group U(2).}

The (3+1)-dimensional Lagrangian then reads
\begin{eqnarray}
  \mathcal L&=&-
  \{\frac{1}{4} F_{\mu\nu}^2+\frac{1}{2}(D_\mu S)^2+\frac{1}{2}(D_\mu P)^2+\frac{1}{2}
  g^2(S\times P)^2\}\nonumber\\
  \label{eq:L4d}
  &&-\{
  \bar\psi\gamma^\mu D_\mu\psi +ig\bar\psi (S\times \psi)
+g\bar\psi\gamma_5(P\times\psi)\}.
\end{eqnarray}

We choose the symmetry-breaking Higgs field as $S^a\equiv A_6^a=v
\delta^a_3$ in the trivial sector. The BPS monopoles are of the
form (for $A_0=0$) \cite{Prasad:1975kr} \bea
A_i^a&=&\epsilon_{aij} {x^j\0g r^2}(1-K(mr)),\\
S^a&=&\delta^a_i{x^i\0g r^2} H(mr), \eea with $H=m r \coth(mr)-1$
and $K=mr/\sinh(mr)$, where $m=gv$ is the mass of the particles
that are charged under the unbroken U(1). The BPS equation
$F_{ij}^a+\epsilon_{ijk}D_k S^a=0$ can be written as a
self-duality equation for $F_{MN}$ with $M,N=1,2,3,6$, and the
classical mass is $M_{\rm cl.}=4\pi m/g^2$.


The susy algebra for the charges $Q^\alpha=\int j^{0\alpha}d^3x$
with $j^A={1\02}\Gamma^B \Gamma^C F_{BC} \Gamma^A \lambda$ reads
\be \{ Q^{\alpha}, \bar Q_{\beta} \}
=-(\gamma^\mu)^\alpha{}_{\beta} P_\mu
+(\gamma_5)^\alpha{}_\beta\, U+i \delta^\alpha_\beta\, V, \quad
\ee with $\alpha,\beta=1,\ldots,4$. In the trivial sector $P_\mu$
acts as $\partial_\mu$, and $U$ multiplies the massive fields by
$m$, but in the topological sector $P_\mu$ are covariant
translations, and $U$ and $V$ are surface integrals.
The classical monopole solution 
saturates the BPS bound $M^2 \ge |\langle U \rangle|^2+|\langle V
\rangle|^2$ by $|U_{\rm cl.}|=M_{\rm cl.}$, and $V_{\rm cl.}=0$.

For obtaining the one-loop quantum corrections, one has to
consider quantum fluctuations about the monopole background. The
bosonic fluctuation equations turn out to be simplest in the
background-covariant Feynman-$R_\xi$ gauge which is obtained by
dimensional reduction of the ordinary background-covariant
Feynman gauge-fixing term in (5+1) dimensions $-{1\02}(D_B[\hat
A]\,a^B)^2$, where $a^B$ comprises the bosonic fluctuations and
$\hat A^B$ the background fields. As has been found in
Refs.~\cite{Kaul:1984bp,Imbimbo:1985mt}, in this gauge the
eigenvalues of the bosonic fluctuation equations (taking into
account Faddeev-Popov fields) and those of the fermionic
fluctuation equations combine such that one can make use of an
index-theorem by Weinberg \cite{Weinberg:1979ma} to determine the
spectral density. This leads to the following (unregularized!)
formula for the one-loop mass correction 
\bea M^{(1)} &=& {4\pi
m_0\0g_0^2} + {\hbar\02} \sum \left( \omega_B-\omega_F \right)\nn
&=&{4\pi m_0\0g_0^2} + {\hbar\02} \int\!{d^3k\0 (2\pi)^{3}}
\sqrt{k^2+m^2}\,\rho_M(k^2),\quad \eea 
with $m_0$ and $g_0$ denoting
bare quantities and 
\be\label{rhoM} \rho_M(k^2)={-8\pi m\0 k^2
(k^2+m^2)}. \ee
This expression is logarithmically divergent and is made finite
by combining it with the one-loop renormalization of $g$, while
$m$ does not need to be renormalized
\cite{Kaul:1984bp,Imbimbo:1985mt}. Combining these two
expressions from the sum over zero point energies and the counter
term, we find that there is a mismatch proportional to
$\epsilon$, but $\epsilon$ multiplies a logarithmically divergent
integral, which in dimensional regularization involves a pole
$\epsilon^{-1}$. We therefore obtain a finite correction of the
form 
\bea\label{M1M} M^{(1)}&=&{4\pi m\0g^2}-\epsilon\times
{2m\0\pi}{\Gamma(-\2-{\epsilon\02}) \0 (2\pi^\2)^\epsilon
\Gamma(-\2)}\nn&&\qquad\qquad \times
\int_0^\infty dk (k^2+m^2)^{-\2+{\epsilon\02}}\nonumber\\
&=&{4\pi m\0g^2}-{2m\0\pi}+O(\epsilon) \eea which because of the
fact that it arises as $0\times\infty$ bears the
hallmark
of an anomaly.

Indeed, as we shall now show, this result is completely analogous
to the case of the $N=1$ susy kink in (1+1) dimensions, where a
nonvanishing quantum correction to the kink mass (in a minimal
renormalization scheme) is associated with an anomaly in the
central charge (which is scheme-independent; in a non-minimal
renormalization scheme there are also non-anomalous quantum
corrections to the central charge).


In Ref.~\cite{Imbimbo:1985mt} it has been argued that in the
renormalization scheme defined above, the one-loop contributions
to the central charge precisely cancel the contribution from the
counterterm in the classical expression. In this parti\-cular
calculation it turns out that the cancelling contributions have
identical form so that the regularization methods of
Ref.~\cite{Imbimbo:1985mt} can be used at least
self-consistently, and also straightforward dimensional
regularization would imply complete cancellations. The result
(\ref{M1M}) would then appear to violate the Bogomolnyi bound.

However, this is just the situation encountered in the
(1+1)-dimensional susy kink. As we have shown in
Ref.~\cite{Rebhan:2002yw} and recapitulated above, 
dimensional regularization gives a
zero result for the correction to the central charge unless the
latter is augmented by the momentum operator in the extra
dimension used to embed the soliton. This is necessary for
manifest supersymmetry, and, indeed, the extra momentum operator
can acquire a nonvanishing expectation value. As it turns out,
the latter is entirely due to nontrivial contributions from the
fermions $\psi={\psi_+\choose\psi_-}$, whose fluctuation
equations have the form \bea
L \psi_+ + i(\6_t+\6_5)\psi_- &=& 0,\\
i(\6_t-\6_5)\psi_+ + L^\dagger \psi_-&=&0. \eea

The fermionic field operator can be written as 
\bea
\psi(x)&=&\int\frac{d^{\epsilon}\ell}{(2\pi)^{\epsilon/2}}
\intsum {d^3k\0(2\pi)^{3/2}}{1\0\sqrt{2\omega}} \nn&&\times\biggl\{ a_{kl}
e^{-i(\omega t - \ell x^5)}
{\sqrt{\omega-\ell}\; \chi_+ \choose - \sqrt{\omega+\ell} \;\chi_- }\nonumber\\
&&\qquad+b_{kl}^\dagger e^{i (\omega t - \ell
x^5)} { \sqrt{\omega-\ell} \;\chi_+ \choose  \sqrt{\omega+\ell}
\;\chi_- } \biggr\}, \eea 
where $\chi_-={1\0\omega_k}L\chi_+$ and
$\chi_+={1\0\omega_k}L^\dagger\chi_-$ with $\omega_k^2=k^2+m^2$,
and the normalization factors $\sqrt{\omega\pm\ell}$ are such that
$L^\dagger L \chi_+ = \omega^2 \chi_+$ and $L L^\dagger \chi_- =
\omega^2 \chi_-$ with $\omega^2=\omega^2_k+\ell^2$. Because of
these normalization factors, one obtains an expression for the
momentum density $\Theta_{05}$ in the extra dimension which has
an even component under reflection in the extra momentum variable
$\ell$ \bea \langle \Theta_{05} \rangle &=&
\int\frac{d^{\epsilon}\ell}{(2\pi)^{\epsilon}} \int
{d^3k\0(2\pi)^3}{\ell\02\omega}
 \Bigl[(\omega-\ell)|\chi_+|^2\nn&&
\qquad\qquad\qquad\qquad\quad+(\omega+\ell)|\chi_-|^2\Bigr]\nonumber\\
&=&\int\frac{d^{\epsilon}\ell}{(2\pi)^{\epsilon}} \int
{d^3k\0(2\pi)^3}\frac{\ell^2}{2\omega}(|\chi_-|^2-|\chi_+|^2) \eea
(omitting zero-mode contributions which do not contribute in
dimensional regularization \cite{Rebhan:2002yw}).

Integration over $x$ then produces the spectral density
(\ref{rhoM}) and finally yields 
\bea\label{deltaZ}
\Delta U_{\rm an}\!&=&\int d^3x\, \langle \Theta_{05} \rangle \nn&=& \int
{d^3k\,d^\epsilon\ell \0 (2\pi)^{3+\epsilon}}
{\ell^2 \0 2 \sqrt{k^2+\ell^2+m^2}}\,\rho_M(k^2) \nonumber\\
&=&\!-4m \int_0^\infty {dk\02\pi} \int\! {d^\epsilon\ell \0
(2\pi)^{\epsilon}}
{\ell^2 \0 (k^2+m^2)\sqrt{k^2+\ell^2+m^2}}\nonumber\\
&=&\!-8 {\Gamma(1-{\epsilon\02})\0(4\pi)^{1+{\epsilon\02}}}
{m^{1+\epsilon}\01+\epsilon} =-{2m\0\pi}+O(\epsilon), \eea which
is indeed equal to the nonzero mass correction obtained above.

This verifies that the BPS bound remains saturated under
quantum corrections, but the quantum corrections to mass and
central charge both contain an anomalous contribution, analogous
to the central-charge anomaly in the 1+1 dimensional minimally
supersymmetric kink.

The nontrivial result (\ref{deltaZ}) is in fact in complete
accordance with the low-energy effective action for $N=2$
super-Yang-Mills theory as obtained by Seiberg and Witten
\cite{Seiberg}.\footnote{We are grateful to Horatiu Nastase for
pointing this out to us.} According to the latter, the low-energy
effective action is fully determined by a prepotential $\mathcal
F(A)$, which to one-loop order is given by \be\label{F1loop}
{\mathcal F_{\rm 1-loop}}(A)=\frac{i}{2\pi}A^{2}\ln
\frac{A^{2}}{\Lambda^2}, \ee where $A$ is a chiral superfield and
$\Lambda$ the scale parameter of the theory generated by
dimensional transmutation. The value of its scalar component $a$
corresponds in our notation to $gv=m$. In the absence of a
$\theta$ parameter, the one-loop renormalized coupling is given
by \be\label{taua} {4\pi i\0g^2}=\tau(a)={\partial^2 {\mathcal F}
\0 \partial a^2} ={i\0\pi}\left(\ln{a^2\0\Lambda^2}+3\right). \ee
This definition agrees with the ``minimal'' renormalization
scheme that we have considered above, because the latter involves
only the zero-momentum limit of the two-point function of the
massless fields. For a single magnetic monopole, the central
charge
is given by \be |U|=a_D={\partial {\mathcal F}\0\partial a}=
{i\0\pi}a\left(\ln{a^2\0\Lambda^2}+1\right) ={4\pi
a\0g^2}-{2a\0\pi}, \ee and since $a=m$, this exactly agrees with
the result of our direct calculation in (\ref{deltaZ}).

Now, the low-energy effective action associated with
(\ref{F1loop}) has been derived from a consistency requirement
with the anomaly of the U(1)$_R$ symmetry of the microscopic
theory. The central-charge anomaly, which we have identified as
being responsible for the entire nonzero correction
(\ref{deltaZ}), is evidently consistent with the former. Just as
in the case of the minimally supersymmetric kink
in 1+1 dimensions, it constitutes a new anomaly\footnote{%
The possibility of central-charge anomalies in $N=2$
super-Yang-Mills theories has most recently also been noted in
\cite{Shifman:2003uh}, however without a calculation of the
coefficients.} that had previously been ignored in direct
calculations \cite{Kaul:1984bp,Imbimbo:1985mt}
of the quantum corrections to 
the $N=2$ monopole.

\newpage

\end{document}